\documentclass[aps,twocolumn,showpacs,preprintnumbers,prx,amsmath,amssymb,amsfonts,superscriptaddress,floatfix]{revtex4-1}
\usepackage[latin1]{inputenc}
\usepackage{graphics}
\usepackage{color}
\usepackage{amssymb}
\usepackage{amsmath}
\usepackage{overpic}
\usepackage{epstopdf}
\usepackage{bm}
\usepackage{graphicx}
\usepackage{subfigure}

\usepackage{xr}
\begin{document}

\title{Sn(\uppercase\expandafter{\romannumeral2})-containing
phosphates as optoelectronic materials}

\author{Qiaoling Xu}
\affiliation{College of Materials Science and Engineering and Key Laboratory of Automobile Materials of MOE, Jilin University, Changchun 130012, China}

\author{Yuwei Li}
\affiliation{Department of Physics and Astronomy, University of Missouri, Columbia, MO 65211-7010 USA}

\author{Lijun Zhang}
\email{Corresponding author: lijun$\_$zhang@jlu.edu.cn}
\affiliation{College of Materials Science and Engineering and Key Laboratory of Automobile Materials of MOE, Jilin University, Changchun 130012, China}

\author{Weitao Zheng}
\affiliation{College of Materials Science and Engineering and Key Laboratory of Automobile Materials of MOE, Jilin University, Changchun 130012, China}

\author{David J. Singh}
\email{Corresponding author: singhdj@missouri.edu}
\affiliation{Department of Physics and Astronomy, University of Missouri, Columbia, MO 65211-7010 USA}
\affiliation{College of Materials Science and Engineering and Key Laboratory of Automobile Materials of MOE, Jilin University, Changchun 130012, China}

\author{Yanming Ma}
\email{Corresponding author: mym@jlu.edu.cn}
\affiliation{State Key Laboratory of Superhard Materials, Jilin University, Changchun 130012, China}

\date{\today}

\begin{abstract} 
We theoretically investigate
Sn(\uppercase\expandafter{\romannumeral2}) phosphates as
optoelectronic materials using first principles calculations. 
We focus on known prototype 
materials Sn$_n$P$_2$O$_{5+n}$ (n=2, 3, 4, 5) and
a previously unreported compound,
SnP$_2$O$_6$ (n=1), 
which we find using global optimization structure prediction.
The electronic structure calculations indicate that these compounds all
have large band gaps above 3.2 eV,
meaning their transparency to visible light.
Several of these compounds show relatively low hole
effective masses ($\sim$2-3 m$_0$), comparable the electron masses.
This suggests potential bipolar conductivity depending on doping.
The dispersive valence band-edges underlying the low hole masses, 
originate from the anti-bonding hybridization between the Sn 5\emph{s} orbitals
and the phosphate groups.
Analysis of structure-property
relationships for the metastable structures
generated during structure search shows
considerable
variation in combinations of band gap and carrier effective masses,
implying chemical tunability of these properties.
The unusual combinations
of relatively high band gap, low carrier masses
and high chemical stability suggests possible optoelectronic applications
of these Sn(II) phosphates, including \emph{p-type} transparent conductors.
Related to this, calculations for doped material indicate low
visible light absorption, combined with high plasma frequencies.
\end{abstract}

\maketitle

\section{\textbf{INTRODUCTION}}

Materials for optoelectronic application must have a suitable
combination of optical properties, including band gap,
and electronic properties, such as mobility and dopability.
A particularly important class of optoelectronic materials are
the transparent conductors. High performance \emph{n-type} transparent
conducting oxides (TCOs) are known, including Sn doped
In$_2$O$_3$ (ITO). The best \emph{p-type} TCOs have inferior
performance due to lower mobility and higher optical absorption
\cite{ginley,kawazoe}.
The difficulties in realizing high-performance \emph{p-type} TCOs
include the rather localized O \emph{p} derived
valence bands of most oxides. The valence bands then have
heavy effective masses and low hole mobilities. Furthermore,
oxide chemistry strongly favors an O$^{2-}$ valence leading
to low energy deep defect levels that prevent high \emph{p-type}
carrier concentrations. These difficulties suggest the exploration
of alternate chemistries to find new optoelectronic materials.

One approach is to take advantage of the
coupling/hybridization between O \emph{p} and states from other atoms, 
as exemplified by delafossite CuAlO$_{2}$.
In that Cu$^{1+}$ compound the valence band edge has substantial
Cu \emph{d} character. The hybridization between the Cu \emph{d} and O \emph{p} states
improves the dispersion of the valence bands,
while the chemical tendency of Cu to favor higher valence
states, e.g. Cu$^{2+}$ enables heavy \emph{p-type} doping
\cite{kawazoe1997p}.
The effective mass of CuAlO$_2$ is reported as $m^*$=2.6 $m_0$,
comparable to other \emph{p-type} TCO materials, e.g. SrCu$_2$O$_2$,
with $m^*$=2.1 $m_0$ and ZnRh$_2$O$_4$, with $m^*$=3.4 $m_0$.
\cite{hautier2013identification,gillen}.
SnO, which has a band gap somewhat to small to be a good TCO
in the visible, has a hole density of states effective mass $m^*$=0.9 $m_0$.
\cite{peng}
Nonetheless, the lower dispersion of 
the valence bands of CuAlO$_2$ relative to the conduction bands of
ITO leads to a large number of bands below the valence band maximum (VBM)
including the 1-3 eV binding energy range. When doped, optical
transitions from these bands to the empty states at the valence
band maximum can degrade the transparency. A similar strategy
has been used with certain non-transition metal elements
where the VBM can be from hybridized metal 
\emph{s} - O \emph{p} combinations.
These include low-valence state cations of group 
\uppercase\expandafter{\romannumeral3},
\uppercase\expandafter{\romannumeral4},
\uppercase\expandafter{\romannumeral5}
metals (for instance Sn$^{+2}$, Pb$^{+2}$, Bi$^{+3}$, Tl$^{+}$, etc.).
An example is the SnO binary compound \cite{ogo2008p}
(n.b., Sn compounds are particularly attractive because Sn is a relatively
low cost, benign ingredient).
Ternary phases have also been studied,
with promissing results both for transparent conducting and electronic
application.
\cite{kim,chambers}
Theoretical work has mainly emphasized finding 
suitable combinations of band gap and hole effective mass
\cite{hautier2013identification}.

Here we explore a different class of materials, specifically 
metal phosphate compounds, using
Sn(\uppercase\expandafter{\romannumeral2}) phosphates as
a first example (here we refer to both ortho- and pyro-phosphates
generically as phosphates). We investigate band gap and effective mass,
as well as the optical properties when doped.
The rationale is that the phosphates
have very stable phosphate polyanions 
that may provide good transport along the anion 
backbone and are based on Zintl chemistry, that may facilitate realization 
of a large number of compounds from which to select and may additionally
provide enough flexibility to enable doping.
The stability of phosphates is key to many of their applications,
including waste storage forms \cite{ewing,day,oelkers},
catalysts \cite{clearfield,banares},
bonding agents in refractory and other ceramics \cite{kingery},
laser glass for very high power lasers \cite{campbell},
battery materials \cite{padhi,yu},
biological processes \cite{westheimer},
and others.
It is also noteworthy that phosphates
can have significant electronic activity, as in catalysis,
energy transport in scintillators, \cite{nakazawa,wojtowicz}
and in biology.
This further motivates the present work.
Specifically, the large phosphate anions that can form strongly
bonded networks suggests the possibility of charge transport. Furthermore,
metal vacancies would be
expected to \emph{p-type} dopants in these compounds, while the very high
chemical stability of the phosphate polyanions could prevent the
formation of compensating defects on the polyanion lattice.
However, we note
that there has to date been very limited investigation of phosphates
as electronic materials \cite{lentz}.

The main valence bands of these phosphates come from the bonding 
states of the phosphate backbone, while the conduction bands come from
metal states that lie in the large gap between
the bonding and antibonding states of the phosphate groups
\cite{singh-pb}.
In the case of Sn(II),
the low valence state of Sn suggests the possibility that
Sn \emph{s} states may also occur at or near the 
VBM, which could provide additional connectivity between 
the phosphate groups and improve \emph{p-type} conduction. As discussed below,
we do
find that the Sn \emph{s} states in these compounds occur at the top of the 
valence bands, so that the VBM has predominant Sn \emph{s} character.

Key initial questions to address are whether Sn(II) phosphates
will have suitable gaps, whether any compounds exist that have
sufficient connection between the phosphate units to enable light
valence band masses and whether other properties are consistent
with \emph{p-type} optoelectronic applications,
especially \emph{p-type} transparent conducting behavior.
It is also of interest to explore the range of properties that can be
realized, as applications will no doubt benefit from chemical tuning
\cite{lentz}.

\begin{figure}
\centerline{\includegraphics[width=3.5in]{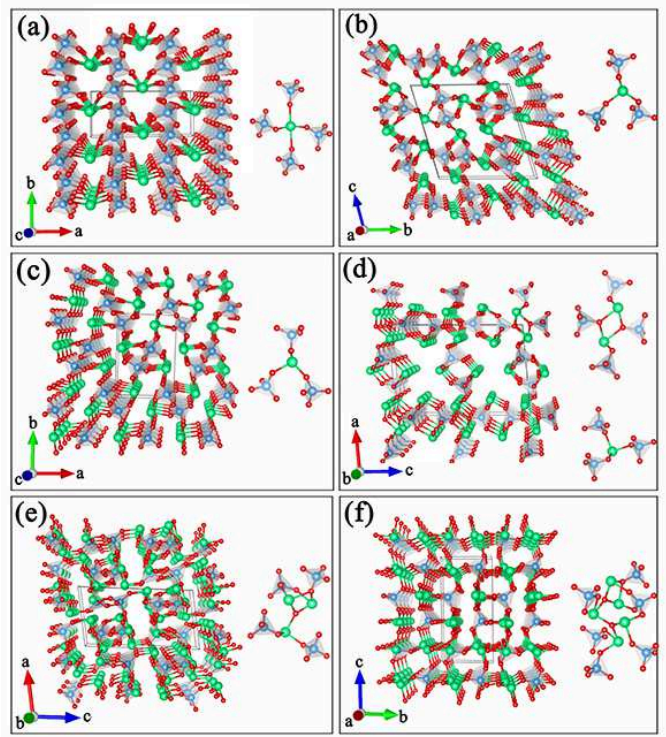}}
\caption{Crystal structures of
Sn$_n$P$_2$O$_{5+n}$ (n=1, 2, 3, 4, 5). (a) is for n=1, predicted stable 
configuration (with the symmetry $Cc$, $Z$=4).
(b, c) is for n=2; the corresponding
structures are the high-temperature $\alpha$-phase (with the $P2_{1}/n$
symmetry, $Z$=2) and the room-temperature $\beta$-phase (with the $P\bar{1}$
symmetry, $Z$=4). (d, e, f) is for n=3 (with $P2_{1}/c$ symmetry, $Z$=4),
n=4 (with the $P2_{1}/n$ symmetry, $Z$=4)
and n=5 (with the symmetry $P\bar{1}$, $Z$=2),
depicting the connections of the basic structural units.
The Sn, P and O atoms are shown in green, blue, and red, respectively.}
\label{structures}
\end{figure}

A series of stoichiometric Sn(II) phosphate compounds 
have been reported experimentally. 
Here we initially focus on a group of  prototype materials, 
Sn$_n$P$_2$O$_{5+n}$ (n=2, 3, 4, 5) (Fig. \ref{structures}).
For n=2, there are two phases: a high-temperature $\alpha$-phase
and a room-temperature $\beta$-phase.
We also performed structure searches for
better connected Sn(\uppercase\expandafter{\romannumeral2}) 
phosphates (n=1),
using the principle of obtaining better connectivity
by choosing a stoichiometry that is 
expected to favor longer pyrophosphate chains.
We note that such pyrophosphates also form glasses, often with properties
similar to the corresponding crystalline phases, which may be advantageous for 
TCO applications, where amorphous phases such as IGZO (In-Ga-Zn-O) are useful
\cite{nomura,kamiya}. The high performance of amorphous
IGZO shows that high degrees of crystalline perfection are not always needed
in TCOs.

\section{\textbf{COMPUTATIONAL METHODS}}

The present calculations were done within density functional
theory (DFT) using the Vienna ab initio software package (VASP)
 \cite{PhysRevB.54.11169}, with projector augmented-wave 
pseudopotentials. 
2\emph{s}$^{2}$2\emph{p}$^{4}$, 5\emph{s}$^{2}$5\emph{p}$^{2}$ and 3\emph{s}$^{2}$3\emph{p}$^{3}$
valence state configurations for O, Sn and P, respectively, were used.
We used the exchange correlation functional of Perdew,
Burke and Ernzerhof (PBE) for the structure determinations
\cite{PhysRevB.46.6671,PhysRevLett.77.3865},
in conjunction with an energy cutoff of 520 eV for the wave-function
expansion and a grid spacing of 2$\pi$$\times$$0.04$ \AA$^{-1}$
for Brillouin zone integration. The structures (including lattice
parameters and internal atomic positions) are fully optimized via 
total energy  minimization. (See supplementary
information, Table S1 which shows the
calculated lattice parameters in relation to experiment.)

The band gap plays a key role in optoelectronic materials.
We used the modified Becke-Johnson (mBJ) potential
\cite{tran2009accurate}
to calculate electronic structures and optical properties.
In contrast to standard density functionals, the mBJ potential
gives band gaps in accord with experiment for most
semiconductors and insulators that do not involve transition
metal \emph{d} or rare earth \emph{f} derived levels
\cite{tran2009accurate,singh-mbj,koller}.
As a test, we compared the electronic structures  obtained with
the mBJ potential to those obtained by the HSE06
\cite{heyd2003hybrid,heyd2006erratum}
hybrid functional method. (See supplementary
information, Fig. S1, which shows the comparison
between the mBJ and HSE06 band structures.
As seen, they give almost identical band gaps and very
similar band dispersion curves.)
We also did tests comparing the electronic structures with
all electron calculations performed with the general
potential linearized augmented plane-wave (LAPW) method
\cite{singh2006planewaves} using the
mBJ \cite{tran2009accurate} potential functional.
We used the BoltzTraP code \cite{madsen2006boltztrap}
to evaluate the effective transport masses for
conductivity. These were determined by
calculating the direction averaged $\sigma/\tau$ at a temperature
of 300 K and the Fermi level set to the band edge. The mass is then
given by the ratio of the number of carriers, $n$, over $\sigma/\tau$
(note that $\sigma/\tau$ is proportional to the square plasma frequency,
i.e. the optical Drude $n/m$,
which is the quantity relevant for electrical transport).
As mentioned, we performed structure prediction
calculations for SnP$_2$O$_{6}$.
These were done based on first principles DFT calculations
combined with the particle swarm optimization algorithm as
implemented in the CALYPSO code
\cite{PhysRevB.82.094116,CPC.183.2063}.
We also particularly calculated the optical absorption spectra of the
compounds when doped \cite{medvedeva,ong,li}.
This is an important but often neglected aspect for identifying
transparent conductors.

\begin{figure}[h]
\centerline{\includegraphics[width=3.5in]{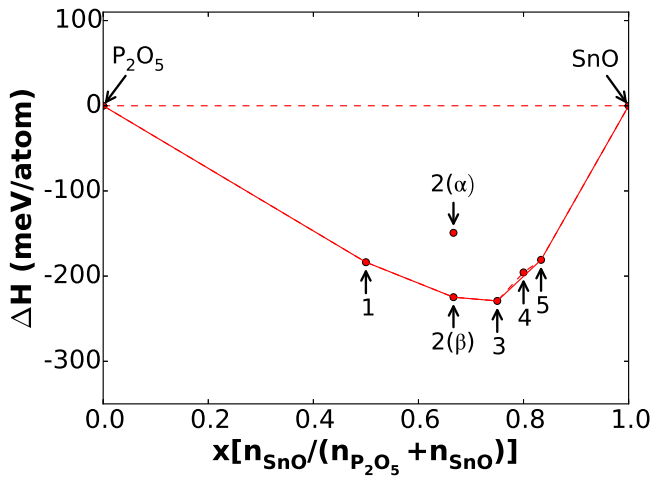}}
\caption{Calculated formation enthalpies $\Delta$H (in
meV/atom), of various Sn(\uppercase\expandafter{\romannumeral2}) 
phosphates with respect to decomposition into the binary compounds,
SnO and P$_2$O$_5$.
Note the large magnitudes of negative $\Delta$H, consistent with the known
generally high stability of metal phosphates.}
\label{convexhull}
\end{figure}

\section{\textbf{RESULTS AND DISCUSSION}}

We begin
with known compounds, Sn$_n$P$_2$O$_{5+n}$ (n=2, 3, 4, 5). As mentioned,
Sn$_2$P$_2$O$_7$ shows two phases with a reversible 
structural transition at 623 K.
These is a high temperature $\alpha$-phase with spacegroup $P2_1/n$
and a $\beta$-phase, spacegroup
$P\bar1$, obtained at room temperature  \cite{Victoria2005}.
Crystalline room-temperature phases of Sn$_3$P$_2$O$_8$ \cite{Mathew1977},
Sn$_4$P$_2$O$_9$ 
and Sn$_5$P$_2$O$_{10}$ \cite{fan2008} have spacegroups
$P2_1/c$, $P2_1/n$ and $P\bar1$, respectively.
Fig. \ref{convexhull} shows their
phase stability with respect to decomposition into the binary compounds
SnO and P$_2$O$_5$, 
{using a convex hull plot}.
Except for $\alpha$-Sn$_2$P$_2$O$_7$,
which is the metastable phase stabilized at high
temperature \cite{Victoria2005}, 
all the materials show clear (n=1, 2($\beta$), 3, 5))
or marginal (n=4) thermodynamic stablity against decomposition.
This is in accord with experiment.
Note that these compounds
have large negative formation enthalpies, 
indicating their high chemical stabilities.
As shown in Fig. \ref{structures}, all of the compounds
consist of PO$_{4}$ tetrahedra (joined by a corner shared O in the
case of the pyrophosphates), and interstitial Sn, coordinated by the O
of the phosphate anions. 
{This type of structure reflects the very high stability of the
phosphate and pyrophosphate anions.}

\begin{figure}[h!]
\centerline{\includegraphics[width=\columnwidth]{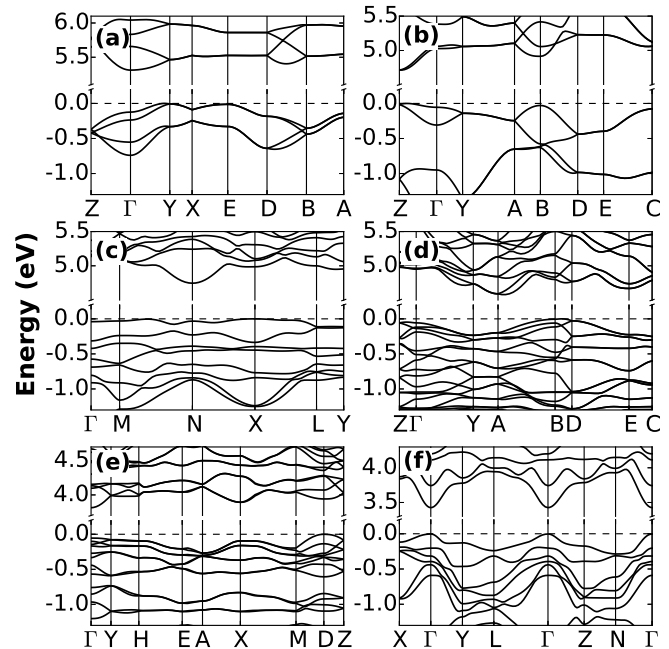}}
\caption{Calculated band structures of Sn$_n$P$_2$O$_{5+n}$
(a (n=1), b (n=2 ($\alpha$)), c (n=2 ($\beta$)), d (n=3), e (n=4), f (n=5))  
using the mBJ potential.}
\label{band}
\end{figure}

\begin{figure}
\centerline{\includegraphics[width=\columnwidth]{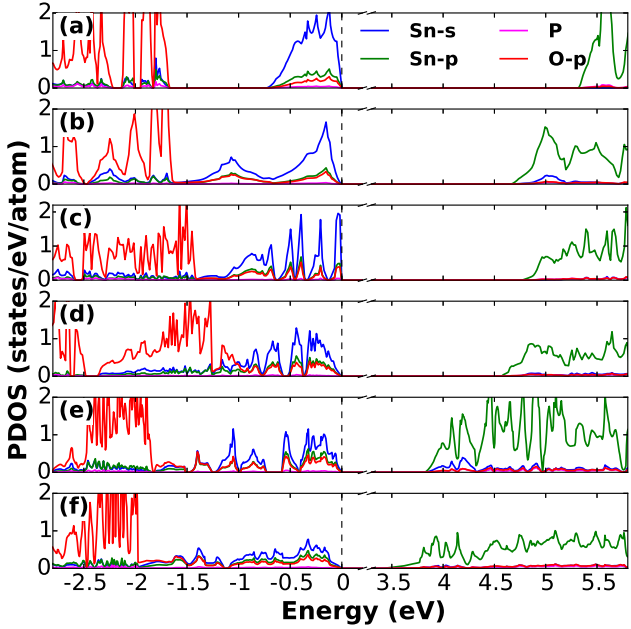}}
\caption{Calculated projected densities of states (PDOS) for Sn$_n$P$_2$O$_{5+n}$
(a (n=1), b (n=2 ($\alpha$)), c (n=2 ($\beta$)), d (n=3), e (n=4), f (n=5)).
The blue, green, magenta, and red lines correspond to Sn \emph{s}, Sn \emph{p}, P (total), and O \emph{p} states, respectively. }
\label{tdos}
\end{figure}

{Band structure plots are given in Fig. }\ref{band}.
{The corresponding electronic densities of states (DOS) are
given in Fig.} \ref{tdos}.
The valence bands are associated with the Sn 5\emph{s} states,
with anti-bonding hybridization of the phosphate groups. 
These Sn 5\emph{s} states occur
at the top of the valence bands, and at the very top the character is
predominantly Sn \emph{s}.
Thus from the point of view of hole transport
the Sn \emph{s} orbitals link
the phosphate groups, which can potentially form conducting channels.

\begin{figure}[h!]
\centerline{\includegraphics[width=3.5in]{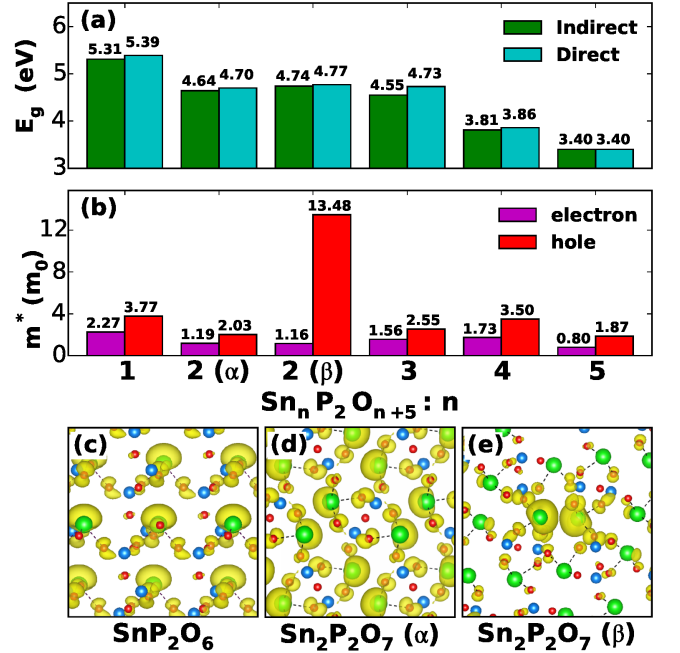}}
\caption{The top panel shows the indirect and 
direct band gap for the n=1, 2 ($\alpha$), 2 ($\beta$), 3, 4 and 5 structures.
Corresponding electron and hole effective masses  are shown in b, respectively. 
Band decomposed charge density of valence-band top with isosurface level 0.002 
for various configurations: n=1, n=2 ($\alpha$) and n=2 ($\beta$), are plotted in the bottom panel.
The color code: Sn green, P blue, O red.}
\label{gap-band-charge}
\end{figure}

\begin{figure}[h!]
\centerline{\includegraphics[width=0.9\columnwidth]{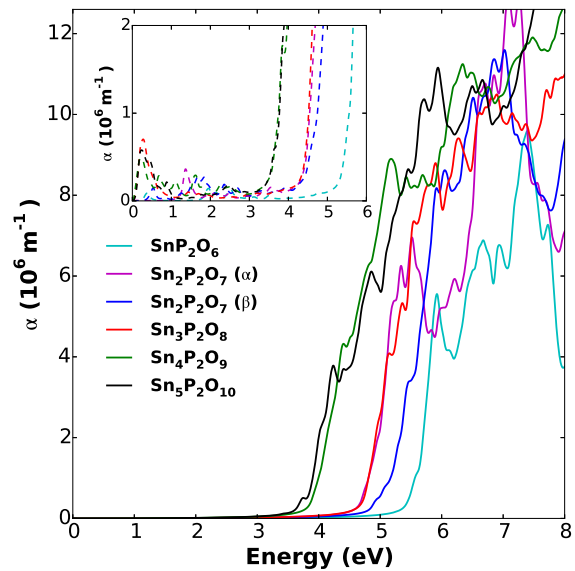}}
\caption{Calculated absorption spectra for different
compounds,  Sn$_n$P$_2$O$_{5+n}$. The color information: 
cyan (n=1), magenta (n=2 ($\alpha$)), blue (n=2 ($\beta$)), red
(n=3), green (n=4), black (n=5). The dashed lines in the inset show
the absorption spectra doping 0.05 holes per Sn. Note the low absorption
in the visible (1.7 eV -- 3.25 eV) even with heavy doping.  }
\label{optics}
\end{figure}

We find that
all  compounds show large band gaps
based on the mBJ calculations, as summarized in Fig. \ref{gap-band-charge}a. 
In particular,
we obtain 5.31 eV for SnP$_2$O$_6$ (Fig. \ref{band}a)
4.64 eV and 4.74 eV for Sn$_{2}$P$_{2}$O$_{7}$ ($\alpha$) and
Sn$_{2}$P$_{2}$O$_{7}$ ($\beta$), respectively  (Fig. \ref{band}b, \ref{band}c).
The values are 4.55 eV, 3.81 eV, and 3.40 eV  for Sn$_{3}$P$_{2}$O$_{8}$,
Sn$_{4}$P$_{2}$O$_{9}$, and Sn$_{5}$P$_{2}$O$_{10}$, respectively
(Fig. \ref{band}d, \ref{band}e, \ref{band}f).
Interestingly, Sn$_{5}$P$_{2}$O$_{10}$ exhibits 
a direct band gap with a relatively small value compared to the
other compounds.
In any case, all these gaps are all well above the visible region
and imply transparency to visible
and in some cases near UV light for undoped materials.
The calculated absorption spectra are shown in Fig. \ref{optics}.

As mentioned, the valence band edges of the five compounds come
from hybridization of Sn \emph{s}
and O \emph{p} states associated with the phosphate anions (Fig. \ref{tdos}).
The conduction bands are Sn \emph{p} derived.
As might be expected, the involvement of Sn \emph{s} at the VBM leads to
dispersive valence bands and thus lowers the effective mass of hole states.
As shown in Fig. \ref{gap-band-charge}b, except for the case
of Sn$_{2}$P$_{2}$O$_{7}$ ($\beta$),
the effective masses of hole states ($\sim$2-3 m$_{0}$) are comparable to 
those of electrons and are comparable to or even lower
than most known \emph{p-type} transparent
conductors. 
We note that this combination of moderate effective mass
for both holes and electrons, high band gap and high chemical stability
is unusual.

The large hole effective mass of
$\beta$ phase Sn$_{2}$P$_{2}$O$_{7}$ comes from the very
flat valence band
adges (Fig. \ref{band}c). The reason is seen in the character of the states
forming the VBM.
We show the band decomposed charge density of the valence-band
edge states for the first three materials in Fig. 
\ref{gap-band-charge}b.
The states around the valence-band edge for 
$\beta$ phase of Sn$_{2}$P$_{2}$O$_{7}$ (Fig. \ref{gap-band-charge}e) come from
Sn \emph{s} orbitals without sufficient
hybridization to the phosphate groups to produce dispersive bands.

In the binary compound, SnO,
the hole effective mass shows strong anisotropy reflecting the
layered structure \cite{ogo2008p}.
This gives rise to strongly anisotropic hole transport behavior.
This is also a characteristic of delafossite CuAlO$_2$. While it is
possible to make and use highly textured films, in general anisotropy
is undesirable for applications.
In the Sn(\uppercase\expandafter{\romannumeral2})
phosphates, the masses are more isotropic as seen from the band structures.

\begin{figure}
\centerline{\includegraphics[width=0.9\columnwidth]{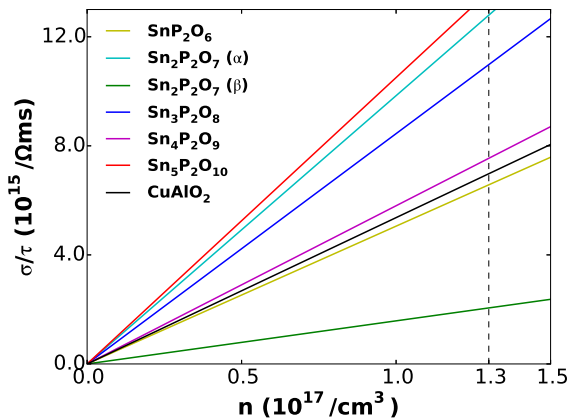}}
\caption{The ratio of optical conductivities $\sigma$ to the
effective inverse scattering rate $\tau$ calculated for n=1, 2($\alpha$),
2($\beta$), 3, 4, 5 structures and CuAlO$_2$, as a function of the carrier
concentration. 
The reported experimental carrier concentration for high performance
CuAlO$_2$ TCO is indicated by the dashed line.}
\label{omiga}
\end{figure}

The calculated $\sigma/\tau$ (proportional to the square plasma 
frequency; $\tau$ is the relaxation time)
as functions of carrier concentration
are given in Fig. \ref{omiga}. For comparison, 
we also show results for a known \emph{p-type} transparent
conductor, delafossite CuAlO$_2$.
Conductivity of $\sim$1 Scm$^{-1}$ has been reported at doping levels of
1.3$\times$10$^{17}$ cm$^{-3}$ for that material.
\cite{kawazoe1997p}
While the scattering rate will be important, we find that considering the
band structure alone, the $\sigma/\tau$ for the compounds follows the
order
$\sigma$ (Sn$_5$P$_2$O$_{10}$) $>$ $\sigma$ (Sn$_2$P$_2$O$_7$ ($\alpha$)) $>$
$\sigma$ (Sn$_3$P$_2$O$_8$) $>$ $\sigma$ (Sn$_4$P$_2$O$_9$) $>$ $\sigma$ (CuAlO$_2$) $>$ 
$\sigma$ (SnP$_2$O$_6$)$>$ $\sigma$ (Sn$_2$P$_2$O$_7$ ($\beta$)), suggesting that
at least room temperature phase Sn$_5$P$_2$O$_{10}$,
Sn$_4$P$_2$O$_9$ and Sn$_3$P$_2$O$_8$,
may be good conductors if doped, and that from an electronic point of view
they may have performance comparable to or better than the delafossite.

We next consider interband transitions in doped material.
This is a key ingredient in the performance of \emph{n-type}
TCOs. To address this important, but often neglected aspect, we
calculate absorption spectra with virtual crystal doping. This allows
transitions from lower bands into the now empty states at the top of
the valence band \cite{ong,li}.
The calculated absorption spectra with
{$p$-$type$}
virtual crystal doping of
0.05 holes per Sn are shown in the inset of Fig. \ref{optics}.
Interestingly, there is an inverse Burstein-Moss shift with doping in
which the apparent optical gap shrinks with hole doping. This reflects
a downward shift of the Sn \emph{p} derived conduction bands when holes are
introduced into the Sn \emph{s} derived valence band. This shift is larger than the
lowering of the Fermi level with hole doping.
Importantly, although there are interband transitions in the visible,
we find that they are very weak, so that high transparency when doped
is possible. This is a key requirement for a transparent conductor, and 
is not met for many materials with otherwise favorable gaps and effective
masses.

An interesting question is whether increasing the length of the
pyrophosphate chains will improve the properties by providing
more connectivity in the polyanion lattice. 
To explore this issue we studied a lower O content composition,
which would lead to longer chains,
specifically, SnP$_2$O$_6$ (see Fig. \ref{structures}a).  
We determined the crystal structure using the particle swarm
optimization algorithms implemented in the CALYPSO code.

We began with a series of structural searches
for the  stoichiometric composition,
SnP$_2$O$_6$. The predicted lowest energy structures
have $Cc$ spacegroups, presented in Fig. \ref{structures}a
(Supplementary information
Table S2 gives the lattice constants and internal atomic coordinates).
{Note that this is a theoretically determined structure based
on density functional calculations. This requires the use of finite
cells, in the present case up to 18 atoms, with some
calculations up to 36 atoms. It is possible that a more
complex larger unit cell structure occurs; in that case the energy
will be lower, and therefore the compound will still be stable.}
The predicted structure has
Sn atoms coordinated by
four O atoms to form square pyramids and
(PO$_3$)$_n$ (n$\geq2$) chains interconnected
by corner shared P$_2$O$_7$ units.
Importantly the result is a connected phosphate network with
interstitial Sn cations.

\begin{figure}[h!]
\includegraphics[width=\columnwidth]{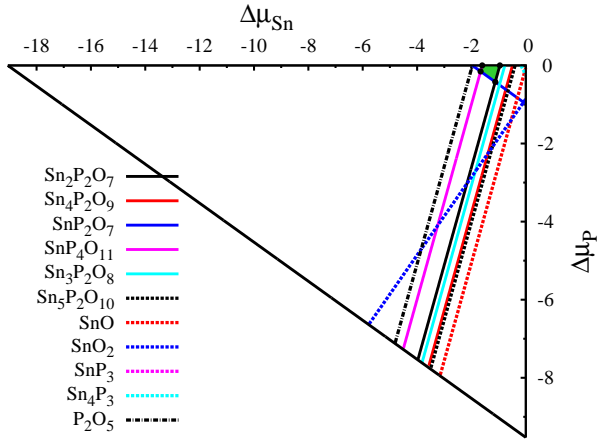}
\caption{Phase stability diagrams of SnP$_2$O$_6$.
Each line represents a known competing phase; the stable region for
SnP$_2$O$_6$ is indicated in green.}
\label{triangle}
\end{figure}

We assessed the
phase stability of the newly predicted SnP$_2$O$_6$ phase by using the standard
thermodynamic stability analysis,
as shown in 
{Fig.}
\ref{triangle}.
The experimentally known potentiallly
competing compounds were considered.
By analyzing the variation in chemical potentials of Sn and P, the predicted
phase shows relatively large stable regions.
{In crystal growth, the thermodynamic equilibrium condition
requires a stable SnP$_2$O$_6$ compound to meet the following three criteria:}

\begin{equation}
\Delta \mu_{Sn} + 2\Delta \mu_{P} + 6\Delta \mu_{O} = \Delta H_{f}(SnP_{2}O_{6}),
\end{equation}

\begin{equation}
\Delta \mu_{I}  \leq 0, (I =Sn,P,O),
\end{equation}

\begin{eqnarray}\nonumber
n_{i}\Delta \mu_{Sn}+m_{i}\Delta \mu_{P} + q_{i}\Delta \mu_{O} \\
\leq \Delta H_{f}(Sn_{n{i}}P_{m_{i}}O_{q{i}}),\ i =1,\dots,Z,    
\end{eqnarray}

\noindent
{where $\Delta \mu_{I} =\mu_{i} - \mu_{i}^{0}$
is the deviation of actual chemical potential of species
$i$ during growth ($\mu_{i}$) from
that of bulk elemental solid or gas phase ($\mu_{i}^{0}$),
and Z refers to the total number of competing phases.
$\Delta H_{f}$ is the enthalpy of formation.
Eq.(1) is for equilibrium growth.
Eq.(2) is the actually allowed range of $\Delta H_{f}$. Eq.(3) is for
energetic stability against competing phases.}

\begin{figure}[h!]
\centerline{\includegraphics[width=\columnwidth]{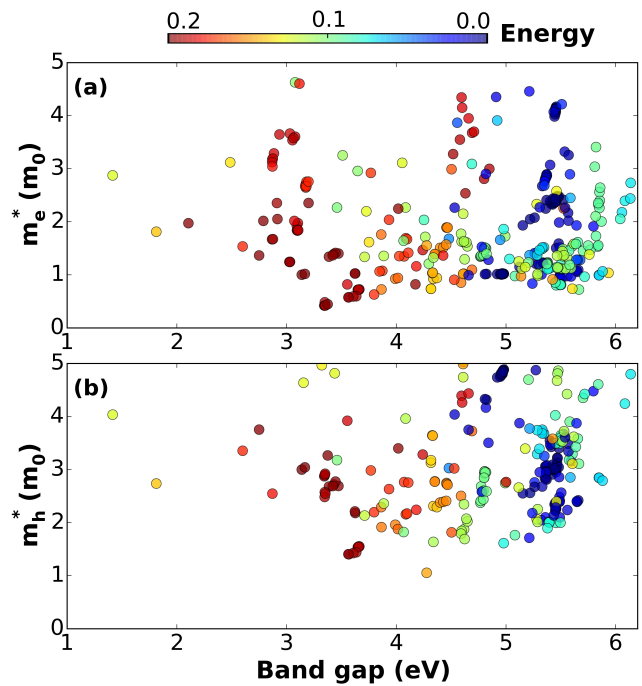}}
\caption{Distribution map of the metastable compounds of
SnP$_2$O$_6$ (n=1) with energies of no more than 0.2 eV/atom higher than 
the ground-state $Cc$ structures, onto  the variables of m$_e$* vs band gap (a)
and m$_h$* \emph{vs} band gaps (b), respectively.
The RGB color bar represents range
of energy from 0.0 eV/atom to 0.2 eV/atom.}
\label{1-mass-Bg}
\end{figure}

Fig. \ref{1-mass-Bg} shows maps of the electron and hole effective mass
\emph{vs} band gap for the lowest energy metastable structures found 
in our structure search.
Remarkably, different from many other families of materials the 
Sn(\uppercase\expandafter{\romannumeral2}) phosphates
do not show a strong connection of mass and band gap. The implication is that 
this chemistry allows for flexible
optimization of both gap and mass in connection
with applications.

We calculated the electronic and optical properties of 
this SnP$_2$O$_6$ phase using the mBJ functional.
Fig. \ref{band}a shows the band structure, which has an
indirect band gap of 5.31 eV 
The absorption spectrum shows a similar value for the
optical gap (see Fig. \ref{optics}).
The band structure shows
a gap from about -1.8 to -0.8 below the VBM.
This gap separates the mainly Sn \emph{s} derived bands making up the 
VBM from lower-lying mainly phosphate derived bands.
We also find that the SnP$_2$O$_6$ structure shows
reasonably low effective mass.
For the $Cc$ phase
the hole and the electron effective masses are
3.77 m$_0$ and 2.27 m$_0$, respectively. 
However, these numbers show that the hole mass of this
more connected structure is not lower than the orthophosphate
Sn$_3$P$_2$O$_8$, suggesting that the density of Sn atoms plays
a more important role than the chain length in these Sn
(\uppercase\expandafter{\romannumeral2}) phosphates.

{We examined the low energy structures in terms of bond valence
sums and other simple metrics, but did not find clear trends for
the effective mass and band gap. We did find that the Sn bond valence
sums for the low energy structures cluster near two and four,
with the lowest energy structures having bond valence near two, as may
be expected from the chemistry of Sn. We also find a common structural
motif of phosphate chain structures, as may be anticipated from the
known chemical stability of such polyanions. The key to our results is
apparently
the chemistry of Sn in a phosphate environment, where the Sn $s$
states occur at the top of the valence band, leading to sizable band gaps
governed by the splitting between the Sn $s$ and Sn $p$ levels.
The hole effective mass is then controlled by the hopping in the
Sn $s$ derived valence bands. This certainly must depend on the
detailed structure and in particular hopping assisted by the
connectivity of the phosphate backbone,
but also is favored by high Sn concentrations, which occur when
the backbone is less connected.}

\section{\textbf{CONCLUSIONS}}

We report electronic structure calculations for some Sn(II) phosphates,
in the context of optoelectronic applications. We find that these materials
have large band gaps and at the same time still can have moderate 
effective masses for both holes and electrons. The bands near 
the valence band maximum in all the
compounds studied have Sn \emph{s} character hybridized with states associated
with the phosphate anions, while the conduction band is Sn \emph{p} derived.
Calculations of optical properties show that interband transitions in
the visible are weak under hole doping. We also find an interesting
inverse Burstein-Moss shift, which can be
understood in terms of the Sn character of
both the states at the valence band maximum and the conduction band minimum.
Analysis of the metastable structures identified in a structure search
suggests that the mass and band gap can be separately tuned in Sn(II)
phosphates, in contrast to most other families of semiconductors.
It will be of interest to determine whether these compounds can be doped,
and also to examine other phosphate compounds as optoelectronic and electronic 
materials.

\section*{ASSOCIATED CONTENT}
\textbf{Supporting Information.}  
{Calculated electronic band structures of $\alpha$-phase
Sn$_2$P$_2$O$_7$ with
different approaches;
Theoretically optimized lattice parameters of Sn$_n$P$_2$O$_{5+n}$
(n=2, 3, 4, 5), compared with available experimental data.  
Detailed structural information for the newly predicted SnP$_2$O$_6$.}

\section*{ACKNOWLEDGMENTS}

L.Z. acknowledges funding support from the Recruitment
Program of Global Youth Experts in China. Some calculations
were performed in the high performance computing center of
Jilin University and on TianHe-1 (A) of the National
Supercomputer Center in Tianjin. We acknowledge funding
support from the National Key Laboratory of Shock Wave and
Detonation Physics, National Natural Science Foundation of
China (No.11534003), and National Key Research and
Development Program of China (No. 2016YFB0201200).

\bibliography{ref}
\end{document}


\title{\textbf{Supplemental Material} for ``Sn(\uppercase\expandafter{\romannumeral2})-containing
phosphates as optoelectronic materials''}

\author{Qiaoling Xu}
\affiliation{College of Materials Science and Engineering and Key Laboratory of Automobile Materials of MOE, Jilin University, Changchun 130012, China}
\author{Yuwei Li}
\affiliation{Department of Physics and Astronomy, University of Missouri, Columbia, MO 65211-7010 USA}
\author{Lijun Zhang}
\email{lijun$_$zhang@jlu.edu.cn}
\affiliation{College of Materials Science and Engineering and Key Laboratory of Automobile Materials of MOE, Jilin University, Changchun 130012, China}
\author{Weitao Zheng}
\affiliation{College of Materials Science and Engineering and Key Laboratory of Automobile Materials of MOE, Jilin University, Changchun 130012, China}
\author{David J. Singh}
\email{singhdj@missouri.edu}
\affiliation{College of Materials Science and Engineering and Key Laboratory of Automobile Materials of MOE, Jilin University, Changchun 130012, China}
\affiliation{Department of Physics and Astronomy, University of Missouri, Columbia, MO 65211-7010 USA}
\author{Yanming Ma}
\email{mym@jlu.edu.cn}
\affiliation{State Key Laboratory of Superhard Materials, Jilin University, Changchun 130012, China}

\maketitle

\clearpage
\vspace*{\fill}
\begin{figure}[h]
\includegraphics[width=5in]{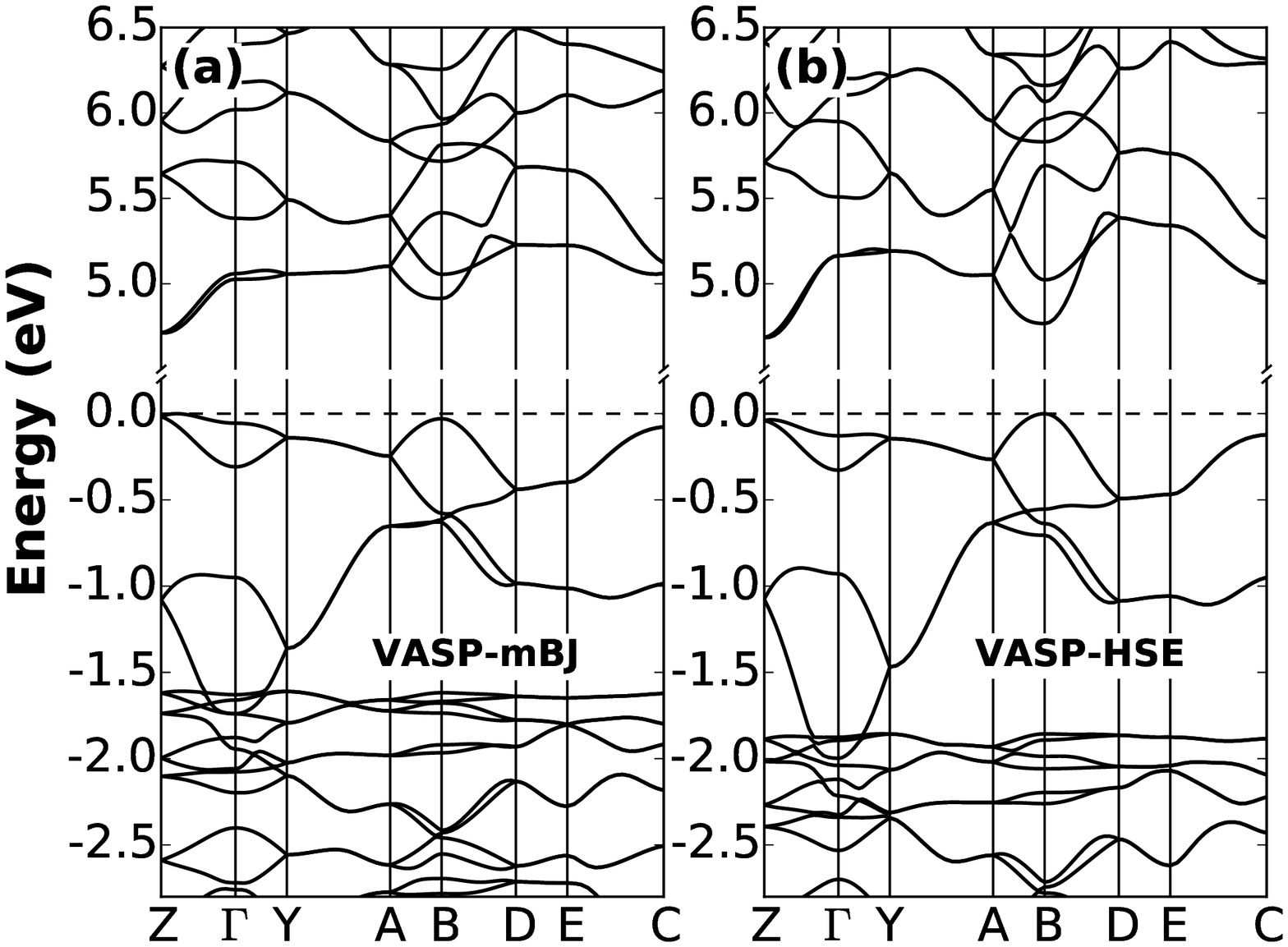}
\centering
\caption{(color online) Calculated electronic band structures of  $\alpha$-
Sn$_2$P$_2$O$_7$ with different functionals: (a)VASP-mBJ, (b)VASP-HSE06.}
\label{2-alphaBand}
\end{figure}
\vfill
\clearpage

\pagebreak
\clearpage

\pagebreak
\clearpage
\begin{table}[h]
\caption{Detailed structural information of predicted stable Sn(PO$_3)_2$.}
\centering
\begin{tabular}{lclccl}
\hline\hline
\textbf{Phase} & \textbf{Lattice parameters} &  \multicolumn{4}{c}{\textbf{Atomic coordinates (fractional)}} \\
 \hline
Sn(PO$_3)_2$-$Cc$  & a=13.139 $\AA$				 & \,\,\,Sn (4a) & \,\,\,   0.626    & \,\,\, 0.116   & \,\,\, 0.999 \\
                                    & b=7.010 $\AA$					 & \,\,\,P \,(4a) & \,\,\,   0.390   & \,\,\, -0.103 & \,\,\, 0.551 \\
                                    & c=10.618 $\AA$				 &\,\,\,P \,(4a) & \,\,\,   0.575    & \,\,\,  0.376  &  \,\,\, 0.372 \\
                                    & $\alpha$=$\gamma$=90$^\circ$        & \,\,\,O \,(4a) & \,\,\,   0.205   & \,\,\,  0.106  &  \,\,\, 0.881 \\
				    & $\beta$=138.896$^\circ$ 			& \,\,\,O \,(4a) & \,\,\,   0.534   & \,\,\,  0.429  &  \,\,\, 0.880 \\
				   &                                 			& \,\,\,O \,(4a) & \,\,\,   0.431   & \,\,\,  -0.308  &  \,\,\, 0.650 \\
				   &                                 		       & \,\,\,O \,(4a) & \,\,\,   0.452   & \,\,\,  -0.057  &  \,\,\, 0.189 \\
				   &                                                 & \,\,\,O \,(4a) & \,\,\,   0.432   & \,\,\,  0.104  &  \,\,\, 0.953  \\
				   &                                                 & \,\,\,O \,(4a) & \,\,\,   0.621   & \,\,\,  0.221  &  \,\,\, 0.505  \\
\hline
\hline
\end{tabular}
\label{structure2}
\end{table}
\vspace{5cm}
\begin{table}[h]
\caption{Optimized lattice parameters (experiment in parenthesis) of the Sn$_n$P$_2$O$_{5+n}$ (n=1-5).}
\centering
\begin{tabular}{lclccl}
\hline\hline
& \textbf{a($\AA$)} & \textbf{b($\AA$)}& \textbf{c($\AA$)} \\
 \hline
n=1 (Sn(PO$_3$)$_2$)				          	  & \,\,\,   13.139    & \,\,\, 7.010  & \,\,\, 10.618 \\
 \hline
 n=2 ($\beta$-Sn$_2$(P$_2$O$_7$))		  	  &  \,\,\,  5.420(5.277)   & \,\,\,11.901(11.541)  & \,\,\, 11.966(11.636) \\
 \hline
 n=2 ($\alpha$-Sn$_2$(P$_2$O$_7$))	          & \,\,\,   7.100(7.176)    & \,\,\, 9.770(9.287)  &  \,\,\, 5.410(5.296) \\
 \hline
n=3 (Sn$_3$(PO$_4$)$_2$)                 		  & \,\,\,   11.457(11.092)   & \,\,\,  4.954(4.830)  &  \,\,\, 16.845(16.405) \\
 \hline
 n=4 (Sn$_4$O(PO$_4$)$_2$)      			  & \,\,\,   7.512(7.353)  & \,\,\,  9.695(9.499)  &  \,\,\, 14.271(13.716) \\
 \hline
n=5 (Sn$_5$O$_2$(PO$_4$)$_2$)       		  & \,\,\,   7.316(7.174)   & \,\,\,  7.354(7.175) &  \,\,\, 13.363(12.895) \\
\hline
\hline
\end{tabular}
\label{structure2}
\end{table}
\clearpage 
